\documentclass[preprintnumbers,showpacs,amsmath,amssymb,twocolumn,floatfix,prd,nofootinbib,showkeys]{revtex4-1}
\usepackage{graphicx}
\usepackage{bbm}
\usepackage{bm}
\usepackage{amsfonts}
\usepackage{epstopdf}
\usepackage[colorlinks,linkcolor=red,anchorcolor=blue,citecolor=green]{hyperref}
\usepackage{subfigure}
\usepackage{float}
\usepackage{enumerate}

\begin{document}
\title{
	Vaidya solution and its generalization in de Rham-Gabadadze-Tolley massive gravity
}
\author{Ping \surname{Li}}
\author{Xin-Zhou \surname{Li}}
\email[]{kychz@shnu.edu.cn}
\author{Xiang-Hua \surname{Zhai}}
\email[]{zhaixh@shnu.edu.cn}
\affiliation{Shanghai United Center for Astrophysics (SUCA), Shanghai Normal University,
100 Guilin Road, Shanghai 200234, China}

\begin{abstract}

We present a detailed study of the Vaidya solution and its generalization in de Rham-Gabadadze-Tolley (dRGT) theory. Since the diffeomorphism invariance can be restored with the St\"{u}ckelberg fields $\phi^a$ introduced, there is a new invariant $I^{ab}=g^{\mu \nu}\partial_\mu \phi^a\partial_\nu \phi^b$ in the massive gravity, which adds to the ones usually encountered in general relativity. There is no conventional Vaidya solution if we choose unitary gauge. In this paper, we obtain three types of self-consistent ansatz with some nonunitary gauge, and find accordingly the Vaidya, generalized Vaidya and furry Vaidya solution. As by-products, we obtain a series of  furry black hole. The Vaidya solution and its generalization in dRGT massive gravity describe the black holes with a variable horizon.
\end{abstract}
\maketitle
\section{Introduction}
It is a significant question whether general relativity (GR) is a solitary theory from both the theoretical and phenomenological sides. One of the modifying gravity theories is the massive deformation of GR. A comprehensive review of massive gravity can be found in \cite{Rham2014}. We can divide the massive gravity theories into two varieties: the Lorentz invariant type (LI) and the Lorentz breaking type (LB). Though for many years it was certain that the theory of LI massive gravity always contains the Boulware-Deser (BD) ghosts \cite{Boulware1972}, a kind of nonlinear extension was recently constructed by de Rham, Gabadadze and Tolley (dRGT) \cite{Gabadadze2009,Rham2010a,Rham2010b,Rham2011a,Zhang2016}. In GR, the spherically symmetric vacuum solution to the Einstein equation is a benchmark, and its massive deformation also plays a crucial role in LI and LB theories. A detailed study of the spherically symmetric solutions is presented in LB massive gravity \cite{Li2016a}, in which we obtain a serviceable formula of the solution to the functional differential equation with spherical symmetry. Using this expression, we give some analytical examples and their phenomenological applications. We present also a detailed study of the black hole solutions in dRGT theory \cite{Li2016b}. Since the diffeomorphism invariance can be restored with the St\"{u}ckelberg fields $\phi^a$ introduced, there is a new invariant $I^{ab}=g^{\mu \nu}\partial_\mu \phi^a\partial_\nu \phi^b$ in the massive gravity, which adds to the ones usually encountered in GR.  In the unitary gauge $\phi^a=x^\mu \delta_\mu ^a$, any inverse metric $g^{\mu\nu}$ that has divergence including the coordinate singularity in GR would exhibit a singularity in the invariant $I^{ab}$. Therefore, there is no conventional Schwarzschild metric if one selects unitary gauge. In the Ref.\cite{Li2016b}, we obtain a self-consistent ans\"{a}tz in the nonunitary gauge, and find that there are seven solutions including the Schwarzschild solution, Reissner-Nordstr\"{o}m solution and five other ones. Furthermore, these solutions may become candidates for black holes in dRGT.

The symmetric tensor field $h_{\mu\nu}\equiv g_{\mu\nu}-\eta_{\mu\nu}$ is the gravitational analogue to the Proca field in the massive electrodynamics, describing all five modes of the massive graviton. With the four St\"{u}ckelberg fields introduced\cite{Hinterbichler2012} and the Minkowski metric replaced by the covariant tensor $\partial_\mu\phi^a\partial_\nu\phi^b\eta_{ab}$, the diffeomorphism invariance can be restored, then the symmetric tensor $H_{\mu\nu}$ describes the covariantized metric perturbation. In the unitary gauge, $H_{\mu\nu}$ reduces to $h_{\mu\nu}$. There is a new basic invariant $I^{ab}=g^{\mu\nu}\partial_\mu\phi^a\partial_\nu\phi^b$ in the massive gravity in addition to the ones usually encountered in GR since the existence of the four scalar fields $\phi^a$. In the unitary gauge, we have $I^{ab}=g^{\mu\nu}\delta_\mu^a\delta_\nu^b$. It is obvious that $I^{ab}$ will exhibit a singularity if $g^{\mu\nu}$ has any divergence including the coordinate singularity for the unitary gauge. De Rham and his colleagues \cite{Berezhiani2012} have pointed out that one would expect the singularities in $I^{ab}$ to be a problem for fluctuations around classical solutions exhibiting it. For this reason, they propose that the solution come true only if $I^{ab}$ is nonsingular. In this paper, we continue to use this conservative rule.

As a corollary of the above point of view, there is no conventional Schwarzschild metric of massive gravity in unitary gauge, which gives rise to the following paradox. According to the vainshtein mechanism \cite{Vainshtein1972}, this solution of massive gravity should approximate the one of GR better and better when we increase the mass of the source. That is to say, this black hole of massive gravity near its horizon should be very similar to that of GR. However, this metric would be singular at the horizon according to the argument above. The Vaidya solution \cite{Vaidya1943,Vaidya1953} is a nonstatic generalization of the Schwarzschild metric in GR. Obviously, there is no conventional Vaidya solution of massive gravity in the unitary gauge. Whether or not there is the conventional Vaidya solution in dRGT with two free parameters is one of the questions that motivates this paper. To find new Vaidya-type solution is another motivation.

Vaidya \cite{Vaidya1943,Vaidya1953} solved Einstein's equations for a spherically symmtric radiating nonrotating body with the energy-momentum tensor of radiation $T^{(rad)}_{\mu\nu}=\rho k_\mu k_\nu$, where $k_\mu$ is a null vector directed radially outward and $\rho$ is defined to be the energy density of the radiation as measured locally by an observer with 4-velocity $v^\mu$, that is to say, $\rho=v^\mu v^\nu T^{(rad)}_{\mu\nu}$. In this work, we study the Vaidya solution and its generalization in dRGT, where two parameters are freely chosen. Furthermore, we release ourselves from the limitation of the unitary gauge $\phi^a=x^\mu \delta_\mu ^a$, and the St\"{u}ckelberg field $\phi^a$ is taken as a "hedgehog" configuration $\phi^i=\phi(u,r)\frac{x^i}r$\cite{Li2016b} and $\phi^0=h(u,r)$, where $u$ is the retarded time \cite{Vaidya1953}. We find a class of Vaidya solutions in dRGT. On the obtained solutions, the singularities in the invariant $I^{ab}$ are absent except for the physical singularity $r=0$, so that these solutions may be regarded as candidates for the dRGT black holes embraced by the radiation.

The paper is organized as follows: Sec. II gives a brief review of dRGT theory \cite{Rham2011a}. In Sec. III, we present three types of self-consistent ansatz with some nonunitary gauge. In Sec. IV, we find the Vaidya solution and a solution of furry black hole under the ansatz I, and in Sec. V the generalized Vaidya solution and the extended solution of furry black hole are found under the ansatz II. The generalized Vaidya solutions are studied under the ansatz III in Sec. VI. The results are summarized and discussed in Sec. VII.

\section{The modified Einstein equations in dRGT theory}
The gravitational action is
\begin{equation}\label{1}
S=\frac{M_{pl}^2}{2}\int d^4x\sqrt{-g}[R+m^2U(g^{\mu\nu},\phi^a)],
\end{equation}
where $R$ is the Ricci scalar, and $U$ is a potential for the graviton which modifies the gravitational sector. The potential is composed of three parts,
\begin{equation}\label{2}
U(g^{\mu\nu},\phi^a)=U_2+\alpha_3U_3+\alpha_4U_4,
\end{equation}
where $\alpha_3$ and $\alpha_4$ are dimensionless parameters, and
\begin{eqnarray}\label{3}
U_2&=&[\mathcal{K}]^2-[\mathcal{K}^2],\nonumber\\
U_3&=&[\mathcal{K}]^3-3[\mathcal{K}][\mathcal{K}^2]+2[\mathcal{K}^3],\\
U_4&=&[\mathcal{K}]^4-6[\mathcal{K}]^2[\mathcal{K}^2]+8[\mathcal{K}][\mathcal{K}^3]+3[\mathcal{K}^2]^2-6[\mathcal{K}^4].\nonumber
\end{eqnarray}
Here the square brackets denote the traces, i.e., $[\mathcal{K}]=\mathcal{K}^\mu_{\ \mu}$ and
\begin{eqnarray}\label{4}
\mathcal{K}^\mu_{\ \nu}&=&\delta^\mu_{\ \nu}-\sqrt{g^{\mu\alpha}\partial_\alpha\phi^a\partial_\nu\phi^b\eta_{ab}}\nonumber\\
&\equiv&\delta^\mu_{\ \nu}-\sqrt{\mathbf{\Sigma}}^\mu_{\ \nu}
\end{eqnarray}
where the matrix square root is $\sqrt{\mathbf{\Sigma}}^\mu_{\ \alpha}\sqrt{\mathbf{\Sigma}}^\alpha_{\ \nu}=\mathbf{\Sigma}^\mu_{\ \nu}$, $g^{\mu\nu}$ is the physical metric, $\eta_{ab}$ is the reference metric and $\phi^a$ are the St\"{u}ckelberg scalars introduced to restore general covariance \cite{Arkani2003}.

Variation of the action with respect to the metric leads to the modified Einstein equations
\begin{equation}\label{5}
G_{\mu\nu}-m^2T^{(\mathcal{K})}_{\mu\nu}=\frac{1}{M_{pl}^2}T^{(rad)}_{\mu\nu},
\end{equation}
where
\begin{equation}\label{6}
T^{(\mathcal{K})}_{\mu\nu}=\frac 1 {\sqrt{-g}}\frac{\delta(\sqrt{-g}U)}{\delta g^{\mu\nu}}.
\end{equation}
From (\ref{4}), we have
\begin{equation}\label{7}
\mathcal{K}^{n\ \mu}_{\ \ \ \ \nu}=\delta^\mu_{\ \nu}+\Sigma_{k=1}^{n}(-1)^k(^n_k)\mathbf{\Sigma}^{\frac{k}{2}\ \mu}_{\ \ \ \ \nu}.
\end{equation}
Thus, $[\mathcal{K}^n]$ can be written as follows,
\begin{equation}\label{8}
\begin{split}
[\mathcal{K}]&=4-[\sqrt{\mathbf{\Sigma}}],\\
[\mathcal{K}^2]&=4-2[\sqrt{\mathbf{\Sigma}}]+[\mathbf{\Sigma}],\\
[\mathcal{K}^3]&=4-3[\sqrt{\mathbf{\Sigma}}]+3[\mathbf{\Sigma}]-[\mathbf{\Sigma}^{\frac{3}{2}}],\\
[\mathcal{K}^4]&=4-4[\sqrt{\mathbf{\Sigma}}]+6[\mathbf{\Sigma}]-4[\mathbf{\Sigma}^{\frac{3}{2}}]+[\mathbf{\Sigma}^2].
\end{split}
\end{equation}

The symmetric tensor $H_{\mu\nu}$ describes the covariantized metric perturbation, which reduces to $h_{\mu\nu}$ in the unitary gauge. Therefore, it is natural to split $\phi^a$ into two parts: $\phi^a=x^a-\pi^a$ and $\pi^a=0$ in the unitary gauge. It is useful that we adopt the following decomposition in the nonunitary gauge,
\begin{equation}\label{9}
\pi^a=\frac{mA^a+\partial^a\pi}{\Lambda^3},
\end{equation}
where $A^a$ describe the helicity $\pm1$, and $\pi$ is the longitudinal mode of the graviton in the decoupling limit \cite{Berezhiani2012}. Moreover, $M_{pl}\rightarrow\infty$ and $m\rightarrow0$ in the decoupling limit \cite{Arkani2003}, while $\Lambda^3\equiv M_{pl}m^2$ is held fixed. This limit represents the approximation in which the energy scale $E$ is much greater than the graviton mass scale.

\section{A self-consistent spherically symmetric ansatz}
\subsection{The metric corresponding to the radiation coordinates}
The front of a gravitational wave (just like that of an electromagnetic wave) provides a unique surface $\Sigma$. Such a null hypersurface $\Sigma$ is described by the equation $x^0=0$ in the radiation coordinate system. The parametric lines of the other coordinates $x^i(i=1,2,3)$ will be situated in $\Sigma$. Thus, there exists a family of noninteracting null hypersurfaces which are described by $x^0=constant$ in this coordinate system. We note that there is a congruence of null geodesics on any null hypersurface $x^0=constant$, which can be used to define a second coordinate $x^1$. Therefore, we should take this congruence as the parametric lines of $x^1$.  In other words, we have $x^2=constant$ and $x^3=constant$ in addition to $x^0=constant$ on each one of the null geodesics of the congruence. Explicitly, the normal vector of surface $\Sigma$, and these geodesics are the parametric lines of $x^1$, so we have $g^{\mu\nu}=\delta^\mu_1$, namely,
\begin{equation}
	g^{\mu\nu}=\left(
	\begin{array}{cccc}
		0&1&0&0\\
		1&g^{11}&g^{12}&g^{13}\\
		0&g^{21}&g^{22}&g^{23}\\
		0&g^{31}&g^{32}&g^{33}
	\end{array}
	\right),
	\end{equation}
and therefore
\begin{equation}
	g_{\mu\nu}=\left(
	\begin{array}{cccc}
		g_{00}&1&g_{02}&g_{03}\\
		1&0&0&0\\
		g_{20}&0&g_{22}&g_{23}\\
		g_{30}&0&g_{32}&g_{33}
	\end{array}
	\right).
	\end{equation}
In the spherical symmetric case, the radiation coordinates $x^\mu$ are usually denoted $(u,r,\theta,\phi)$, where $r$ is the usual radial coordinate and $\theta,\phi$ are generalized polar angles. Accordingly, the general form of the covariant components of metric
\begin{equation}
g_{\mu\nu}=\left(
\begin{array}{cccc}
b^2(u,r)&1&0&0\\
1&0&0&0\\
0&0&-r^2&0\\
0&0&0&-r^2\sin^2\theta
\end{array}
\right),
\label{co-metric}
\end{equation}
and therefore
\begin{equation}\label{13}
g^{\mu\nu}=\left(
\begin{array}{cccc}
0&1&0&0\\
1&-b^2(u,r)&0&0\\
0&0&-r^{-2}&0\\
0&0&0&-r^{-2}\csc^2\theta
\end{array}
\right).
\end{equation}

For the static line element
\begin{equation}
ds^2=b^2(r)du^2+2dudr-r^2d\Omega^2\nonumber
\end{equation}
where $u$ can be interpreted as retarded time coordinate and
\begin{equation}\label{14}
u=t-\int^r_{r_0}\frac{dr}{b^2(r)}.
\end{equation}
Hence we obtain for the null hypersurfaces
\begin{equation}
t-\int^r_{r_0}\frac{dr}{b^2(r)}=constant.
\end{equation}
In the case of the Schwarzschild solution,
\begin{equation}
ds^2=\left(1-\frac{r_s}r\right)du^2+2dudr-r^2d\Omega^2,
\end{equation}
and
\begin{equation}
u=t-r-r_s\ln(r-r_s),
\end{equation}
where $r_s$ is the Schwarzschild radius. Especially, $r_s=0$ and we have $u=t-r$, and the Minkowskian metric
\begin{equation}
ds^2=du^2+2dudr-r^2d\Omega^2.
\end{equation}

From (\ref{co-metric}) we have the Christoffel symbols of the second kind,
\begin{eqnarray}
\Gamma^0_{00}&=&-bb', \Gamma^0_{22}=r, \Gamma^0_{33}=r\sin^2\theta, \Gamma^1_{00}=b\dot{b}+b^3b',\nonumber\\
\Gamma^1_{01}&=&\Gamma^1_{10}=bb', \Gamma^1_{22}=-b^2r, \Gamma^1_{33}=-b^2r\sin^2\theta,\nonumber\\
\Gamma^2_{21}&=&\Gamma^2_{12}=r^{-1}, \Gamma^2_{33}=-\sin\theta\cos\theta, \Gamma^3_{31}=\Gamma^3_{13}=r^{-1}, \nonumber\\
\Gamma^3_{32}&=&\Gamma^3_{23}=\cot\theta.
\end{eqnarray}
All other symbols vanish. The Ricci tensor in radiation coordinates is consequently given by
\begin{eqnarray}\label{20}
R_{00}&=&b^2b'^2+b^3b''+\frac 2 r (b\dot{b}+b^3b'),\nonumber\\
R_{22}&=&\frac{R_{33}}{\sin^2\theta}=-b^2-2bb'r+1,\nonumber\\
R_{01}&=&R_{10}=b'^2+bb''+\frac{2bb'}r,
\end{eqnarray}
where $\dot{b}=\partial b/\partial u, b'=\partial b/\partial r$, and all other components are zero. A straightforward calculation then shows that the Ricci scalar is given by
\begin{equation}\label{21}
R=2\left( b'^2+bb''+\frac{4bb'}r+\frac{b^2-1}{r^2}\right).
\end{equation}
The nonvanishing components of the mixed Einstein tensor $G_\mu^{\ \nu}$ are then given in the following
\begin{eqnarray}\label{22}
G_0^{\ 0}&=&G_1^{\ 1}=-\frac{2bb'}r-\frac{b^2-1}{r^2},\nonumber\\
G_2^{\ 2}&=&G_3^{\ 3}=-\left( b'^2+bb''+\frac{2bb'}r\right).
\end{eqnarray}

\subsection{The ansatz for St\"{u}ckelberg field}

We consider the general form of spherically symmetric ansatz for St\"{u}ckelberg field as follows
\begin{eqnarray}
\phi^0&=&h(u,r),\nonumber\\
\phi^i&=&\phi(u,r)\frac{x^i}r.
\label{ansatz}
\end{eqnarray}
The ansatz (\ref{ansatz}) contains two additional functions $h(u,r)$ and $\phi(u,r)$, which reduces to unitary gauge only if $h(u,r)=u+\int^r_{r_0}\frac{dr}{b^2(r)}$ and $\phi(u,r)=r$ in the static case. The self-consistency of ansatz (\ref{ansatz}) imposes restrictions on $h(u,r)$ and $\phi(u,r)$. Under the ansatz (\ref{ansatz}), the matrix $\mathbf{\Sigma}=(\Sigma^{\mu}_{\ \nu})$ takes the form
\begin{widetext}
\begin{equation}
	\mathbf{\Sigma}=\left(
	\begin{array}{cccc}
		\dot{h}h'-\dot{\phi}\phi'&h'^2-\phi'^2&0&0\\
		(\dot{h}^2-\dot{\phi}^2)-b^2(\dot{h}h'-\dot{\phi}\phi')&(\dot{h}h'-\dot{\phi}\phi')-b^2(h'^2-\phi'^2)&0&0\\
		0&0&\frac{\phi^2}{r^2}&0\\
		0&0&0&\frac{\phi^2}{r^2}
	\end{array}
	\right)
	\end{equation}
\end{widetext}
where dots and primes denote derivatives with respect to $u$ and $r$, respectively.

For a $2\times2$ matrix $\mathbf{M}$, the Cayley-Hamilton theorem tells us that
\begin{equation}
[\mathbf{M}]\mathbf{M}=\mathbf{M}^2+(\det \mathbf{M})\mathbf{I}_2,
\end{equation}
where $\mathbf{I}_2$ is $2\times 2$ identity matrix. We define $\mathbf{\Sigma}_2$ as the upper left-hand $2\times 2$ submatrix of $\mathbf{\Sigma}$ and use $\det \mathbf{M}^n=(\det \mathbf{M})^n$ to find the square root of $\mathbf{\Sigma}_2$,
\begin{equation}
\sqrt{\mathbf{\Sigma}_2}=\frac 1{[\sqrt{\mathbf{\Sigma}_2}]}\left(\mathbf{\Sigma}_2+\sqrt{\det\mathbf{\Sigma}_2}\mathbf{I}_2\right),
\label{sigma2}
\end{equation}
where
\begin{equation}
\det\mathbf{\Sigma}_2=(\dot{h}\phi'-h'\dot{\phi})^2,
\label{det-sigma2}
\end{equation}
and
\begin{equation}
[\sqrt{\mathbf{\Sigma}_2}]=\sqrt{[\mathbf{\Sigma}_2]+2\sqrt{\det\mathbf{\Sigma}_2}}.
\label{bra-sigma2}
\end{equation}
Using (\ref{sigma2})-(\ref{bra-sigma2}), we obtain the recursion formula as follows
\begin{equation}
\mathbf{\Sigma}^{\frac{k+1}2}_2=[\sqrt{\mathbf{\Sigma}_2}]\mathbf{\Sigma}^{\frac k 2}_2-(\det\sqrt{\mathbf{\Sigma}_2})\mathbf{\Sigma}^{\frac{k-1}2}_2,\hspace{0.5cm}(k\geq 1),
\end{equation}
and $\mathbf{\Sigma}^0_2\equiv \mathbf{I}_2$. Thus, we have
\begin{equation}
\mathbf{\Sigma}^{\frac k 2}=\left(
\begin{array}{cc}
      \mathbf{\Sigma}^{\frac k 2}_2&0\\
      0&\left(\frac{\phi}r\right)^k \mathbf{I}_2
      \end{array}
      \right),
      \end{equation}
and
\begin{equation}\label{31}
[\mathcal{K}^n]=4+\Sigma^n_{k=1}(-1)^k\left(\begin{array}{c}n\\k\end{array}\right)\left(\mathbf{\Sigma}^{\frac k 2}_2+2\left(\frac {\phi}r\right)^k\right).
\end{equation}
From (\ref{6}),(\ref{8}) and (\ref{31}), we obtain the non-zero components of $T^{(\mathcal{K})\mu}_{\hspace{0.6cm}\nu}$ as follows
\begin{widetext}
\begin{eqnarray}\label{32}
T^{(\mathcal{K})0}_{\hspace{0.6cm}0}&=&\left(1-\frac{2\phi}r\right)[\sqrt{\mathbf{\Sigma}_2}]+\frac{2\phi}r+\sqrt{\det\mathbf{\Sigma}_2}-\frac{\phi^2}{r^2}+\frac{4}{[\sqrt{\mathbf{\Sigma}_2}]}
\left(\frac{\phi}r-1\right)\left(\dot{h}h'-\dot{\phi}\phi'+\sqrt{\det\mathbf{\Sigma}_2}\right)\nonumber\\
&+&3\alpha_3\left(\left(\frac{\phi^2}{r^2}-1\right)[\sqrt{\mathbf{\Sigma}_2}]+2(1+\sqrt{\det\mathbf{\Sigma}_2})\left(1-\frac{\phi}r\right)-\frac2{[\sqrt{\mathbf{\Sigma}_2}]}
\left(\frac{\phi}r-1\right)^2(\dot{h}h'-\dot{\phi}\phi'+\sqrt{\det\mathbf{\Sigma}_2})\right)\nonumber\\
&-&12\alpha_4\left(\frac{\phi}r-1\right)^2\left([\sqrt{\mathbf{\Sigma}_2}]-(1+\sqrt{\det\mathbf{\Sigma}_2})\right),
\end{eqnarray}
\begin{eqnarray}
T^{(\mathcal{K})0}_{\hspace{0.6cm}1}&=&\frac 2{[\sqrt{\mathbf{\Sigma}_2}]}(h'^2-\phi'^2)\left(\frac{\phi}r-1\right)\left(2-3\alpha_3\left(\frac{\phi}r-1\right)\right),\\
T^{(\mathcal{K})1}_{\hspace{0.6cm}0}&=&\frac 2{[\sqrt{\mathbf{\Sigma}_2}]}(\dot{h}^2-\dot{\phi}^2+b^2(\dot{\phi}\phi'-\dot{h}h'))\left(\frac{\phi}r-1\right)\left(2-3\alpha_3\left(\frac{\phi}r-1\right)\right),\\
T^{(\mathcal{K})1}_{\hspace{0.6cm}1}&=&T^{(\mathcal{K})0}_0+\frac 2{[\sqrt{\mathbf{\Sigma}_2}]}b^2(\phi'^2-h'^2)\left(\frac{\phi}r-1\right)\left(2-3\alpha_3\left(\frac{\phi}r-1\right)\right),\\
T^{(\mathcal{K})2}_{\hspace{0.6cm}2}&=&[\sqrt{\mathbf{\Sigma}_2}]-\frac{2\phi}r-\sqrt{\det\mathbf{\Sigma}_2}+\frac{\phi^2}{r^2}-3\alpha_3\left(\frac{\phi}r-1\right)^2(\sqrt{\mathbf{\Sigma}_2}-2)
-12\alpha_4\left(\frac{\phi}r-1\right)^2
\left([\sqrt{\mathbf{\Sigma}_2}]-(1+\sqrt{\det\mathbf{\Sigma}_2})\right),\label{36}\\
T^{(\mathcal{K})3}_{\hspace{0.6cm}3}&=&T^{(\mathcal{K})2}_{\hspace{0.6cm}2}.
\end{eqnarray}
\end{widetext}
From the modified Einstein equation (\ref{5}) in vacuum, we require that $T^{(\mathcal{K})0}_{\hspace{0.6cm}1}$ and $T^{(\mathcal{K})1}_{\hspace{0.6cm}0}$ vanish which is a self-consistent requisition for the ansatz (\ref{ansatz}). Therefore, the self-consistent ansatz can be classified into three types as follows

Ansatz I:
\begin{equation}
 \begin{split}ds^2&=b^2(u,r)du^2+2dudr-r^2d\Omega^2,\\\phi^0&=h(u,r),\\
\phi^i&=x^i;\end{split}
\end{equation}

Ansatz II:
\begin{equation}
 \begin{split}ds^2&=b^2(u,r)du^2+2dudr-r^2d\Omega^2,\\\phi^0&=h(u,r),\\
\phi^i&=\left(\frac 2{3\alpha_3}+1\right)x^i;\end{split}
\end{equation}

Ansatz III:
\begin{equation}
 \begin{split}ds^2&=b^2(u,r)du^2+2dudr-r^2d\Omega^2,\\\phi^0&=h(u,r),\\
\phi^i&=h(u,r)\frac{x^i}r.\end{split}
\end{equation}
It is easy to verify that $T^{(\mathcal{K})0}_0=T^{(\mathcal{K})1}_1$ under all types. On the other hand, the energy-momentum tensor of a radiating field $T^{(rad)}_{\mu\nu}$ can be described as the geometrical optics form \cite{Carmeli1977}
\begin{equation}\label{41}
T^{(rad)}_{\mu\nu}=-\frac 2{r^2}q(u)\delta^0_\mu\delta^0_\nu.
\end{equation}
Combing now (\ref{13}) and (\ref{41}), we find
\begin{equation}\label{42}
T^{(rad)\mu}_{\hspace{0.8cm}\nu}=0.
\end{equation}

\subsection{The equation between $T^{(\mathcal{K})0}_{\hspace{0.6cm}0}$ and $T^{(\mathcal{K})2}_{\hspace{0.6cm}2}$}

From (\ref{22}), (\ref{32}), (\ref{36}) and (\ref{42}), the modified Einstein equation with and without the radiating field can be rewritten as
\begin{eqnarray}
\frac{(b^2)'}r+\frac{b^2-1}{r^2}&=&m^2T^{(\mathcal{K})0}_{\hspace{0.6cm}0},\label{43}\\
\frac{(b^2)^{''}}2+\frac{(b^2)'}r&=&m^2T^{(\mathcal{K})2}_{\hspace{0.6cm}2}.\label{44}
\end{eqnarray}

There is a mathematical identify relation
\begin{equation}\label{45}
\begin{split}
&\left(\frac{(b^2)'}r+\frac{b^2-1}{r^2}\right)'\\
&=\frac 2 r\left(\left(\frac{(b^2)''}2+\frac{(b^2)'}r\right)-\left(\frac{(b^2)'}r+\frac{b^2-1}{r^2}\right)\right),
\end{split}
\end{equation}
which is the key to the analytical solution. Combing (\ref{43}), (\ref{44}) and (\ref{45}), we obtain
\begin{equation}\label{46}
\left(T^{(\mathcal{K})0}_{\hspace{0.6cm}0}\right)'=\frac 2 r\left(T^{(\mathcal{K})2}_{\hspace{0.6cm}2}-T^{(\mathcal{K})0}_{\hspace{0.6cm}0}\right),
\end{equation}
which is a necessary condition of $T^{(\mathcal{K})\mu}_{\hspace{0.6cm}\nu}$. In general, $T^{(\mathcal{K})0}_{\hspace{0.6cm}0}$ and $T^{(\mathcal{K})2}_{\hspace{0.6cm}2}$ are functions of $b^2(u,r)$ and $h(u,r)$ under three types of ansatz. Under some suitable boundary conditions, there is always a numerical solution to the system composed of two equations (\ref{43}) and (\ref{46}) with two unknown functions. However, the motivation of our work is to find possible exact solutions, so we will settle these types one by one.

\section{solutions under the ansatz I}

In this section, we present a detailed study of solutions under the self-consistent ansatz I in dRGT with two free parameters $\alpha_3$ and $\alpha_4$. The obtained solutions are free of singularities except for the conventional one appearing in GR (for instance, the singularity $r=0$ in the spherically  symmetric solutions).

For the ansatz I, (\ref{32}) and (\ref{36}) can be reduced to
\begin{equation}
T^{(\mathcal{K})0}_{\hspace{0.6cm}0}=-T^{(\mathcal{K})2}_{\hspace{0.6cm}2}=-[\sqrt{\mathbf{\Sigma}_2}]+\sqrt{\det\mathbf{\Sigma}_2}+1,
\end{equation}
where
\begin{equation}
\det\mathbf{\Sigma}_2=\dot{h}^2,
\end{equation}
and
\begin{equation}
[\sqrt{\mathbf{\Sigma}_2}]^2=2\dot{h}(h'+1)-b^2(h'^2-1).
\end{equation}
Thus, (\ref{46}) becomes
\begin{equation}
\left(T^{(\mathcal{K})0}_{\hspace{0.6cm}0}\right)'=-\frac 4 r T^{(\mathcal{K})0}_{\hspace{0.6cm}0},
\end{equation}
which is a separable equation and
\begin{equation}\label{51}
T^{(\mathcal{K})0}_{\hspace{0.6cm}0}=\frac{S(u)}{r^4}.
\end{equation}
Substituting (\ref{51}) into (\ref{43}), we have
\begin{equation}
r(b^2)'+(b^2-1)=\frac{m^2S(u)}{r^2},
\end{equation}
and
\begin{equation}\label{53}
b^2=1-\frac{r_s(u)}r-\frac{m^2S(u)}{r^2}.
\end{equation}

On the other hand, we have the equation of $h(u,r)$ as follows
\begin{equation}\label{54}
-\left(\sqrt{2\dot{h}(h'+1)-b^2(h'^2-1)}\right)+\dot{h}+1=\frac{S(u)}{r^4},
\end{equation}
from which the function $h(u,r)$ can be determined. There exist two cases that (\ref{54}) degenerates and becomes an ordinary differential equation: (i)$h'^2=1$ and $S(u)=0$; (ii)$\dot{h}$, $S(u)$ and $r_s(u)$ are all constants. In reality, case (i) corresponds with the Vaidya solution \cite{Vaidya1953} and case (ii) correlates closely with the solution of furry black hole \cite{Li2016b}.

\subsection{The Vaidya solution in dRGT}

For the case of $h'^2=1$ and $S(u)=0$, (\ref{54}) is reduced to
\begin{equation}
\dot{h}-2\dot{h}^{\frac 1 2}+1=0,\hspace{0.5cm}for\hspace{0.5cm} h'=1,
\end{equation}
or
\begin{equation}
\dot{h}+1=0, \hspace{0.5cm}for\hspace{0.5cm} h'=-1.
\end{equation}
Thus, we obtain
\begin{equation}
h=\pm(u+r).
\end{equation}
In the meantime, (\ref{53}) becomes
\begin{equation}\label{58}
b^2=1-\frac{r_s(u)}r.
\end{equation}
Substituting (\ref{58}) into (\ref{21}), we have the Ricci scalar $R=0$. Since the Ricci scalar and $T^{(\mathcal{K})}_{\mu\nu}$ vanish, the modified Einstein equation may also read as
\begin{equation}
R_{\mu\nu}=-\frac 2{r^2}q(u)\delta^0_\mu \delta^0_\nu.
\end{equation}
From (\ref{20}), we have
\begin{equation}
R_{\mu\nu}=-\frac {\dot{r}_s(u)}{r^2}\delta^0_\mu \delta^0_\nu,
\end{equation}
and
\begin{equation}
q(u)=\frac{d\dot{r}_s(u)}{du}.
\end{equation}

Finally, the Vaidya solution can be written as
\begin{equation}\label{62}
\begin{split}
b^2&=1-\frac{r_s(u)}r,\\
\phi^0&=\pm(u+r),\\
\phi^i&=x^i.
\end{split}
\end{equation}
Due to the existence of the St\"{u}ckelberg field, there is a new basic invariant $I^{ab}=g^{\mu\nu}\partial_\mu\phi^a\partial_\nu\phi^b$ in the massive gravity in addition to the ones usually encountered in GR. de Rham and his colleagues have pointed out that the solution comes true only if $I^{ab}$ is nonsingular \cite{Berezhiani2012}. For the Vaidya solution (\ref{62}), we have
\begin{equation}
\begin{split}
I^{00}&=2-b^2,\\
I^{0i}&=(2-b^2)n^i,\\
I^{ij}&=(n^in^j-\delta^{ij})\frac 1{r^2}-b^2n^in^j.
\end{split}
\end{equation}

\subsection{Furry black hole}

For the case of $\dot{h}=0$ and $S(u)=S$, $r_s(u)=r_s$ ($S$ and $r_s$ are constants), (\ref{54}) is reduced to
\begin{equation}
(1-h'^2)b^2=\left(\frac S{r^4}-1\right)^2,
\end{equation}
and
\begin{equation}
h=\pm\int\left(1-\frac{\left(\frac S{r^4}-1\right)^2}{b^2}\right)^{\frac 1 2}dr,
\end{equation}
where
\begin{equation}
b^2(r)=1-\frac {r_s}r-\frac{m^2S}{r^2}.\nonumber
\end{equation}
This solution is also free of singularities except for the conventional one appearing in GR. In fact, we have
\begin{equation}\label{66}
\begin{split}
I^{00}&=-b^2+\left(\frac S {r^4}-1\right)^2,\\
I^{0i}&=\mp b^2\left(1-\frac 1 {b^2}\left(\frac S{r^4}-1\right)^2\right)^{\frac 1 2}n^i,\\
I^{ij}&=(n^in^j-\delta^{ij})\frac 1{r^2}-b^2n^in^j.
\end{split}
\end{equation}
Using the coordinate transformation (\ref{14}), we obtain the furry black hole solution in the Schwarzschild coordinate
\begin{equation}
\begin{split}
ds^2&=\left(1-\frac{r_s}r-\frac{m^2S}{r^2}\right)dt^2\\
&-\left(1-\frac {r_s}r-\frac{m^2S}{r^2}\right)^{-1}dr^2-r^2d\Omega^2,\\
\phi^0&=\pm\int\left(\frac{\frac{S^2}{r^6}-S(\frac 2 {r^2}-m^2)+r_sr}{m^2S+r_sr-r^2}\right)^{\frac 1 2}dr,\\
\phi^i&=x^i.
\end{split}
\end{equation}

\section{solutions under the ansatz II}

In this section, we find out a generalized Vaidya solution and extended furry black holes under self-consistent ansatz II in dRGT with two free parameters $\alpha_3$and $\alpha_4$.

For the ansatz II, (\ref{32}) and (\ref{36}) can be reduced to
\begin{equation}
T^{(\mathcal{K})0}_{\hspace{0.6cm}0}=\left(3-\frac{16\alpha_4}{3\alpha_3^2}\right)([\sqrt{\mathbf{\Sigma}_2}]-\sqrt{\det\mathbf{\Sigma}_2})+\frac 4{9\alpha_3^2}(12\alpha_4-1)-3,
\end{equation}
and
\begin{equation}
T^{(\mathcal{K})2}_{\hspace{0.6cm}2}=T^{(\mathcal{K})0}_{\hspace{0.6cm}0}-2([\sqrt{\mathbf{\Sigma}_2}]-\sqrt{\det\mathbf{\Sigma}_2})-\frac 8 {3\alpha_3}[\sqrt{\mathbf{\Sigma}_2}]+2\left(\frac 2 {3\alpha_3}+1\right)^2,
\end{equation}
where
\begin{equation}
\det\mathbf{\Sigma}_2=\left(\frac 2{3\alpha_3}+1\right)^2\dot{h}^2,
\end{equation}
and
\begin{equation}
[\sqrt{\mathbf{\Sigma}_2}]^2=2\dot{h}h'-b^2\left(h'^2-\left(\frac 2{3\alpha_3}+1\right)^2\right)+2\sqrt{\det\mathbf{\Sigma}_2}.
\end{equation}
$T^{(\mathcal{K})2}_{\hspace{0.6cm}2}$ is clearly a linear function of $T^{(\mathcal{K})0}_{\hspace{0.6cm}0}$ if $[\mathbf{\Sigma}_2]$ or $\sqrt{\det\mathbf{\Sigma}_2}$ is constant. In this case, we have
\begin{equation}\label{72}
T^{(\mathcal{K})2}_{\hspace{0.6cm}2}-T^{(\mathcal{K})0}_{\hspace{0.6cm}0}=-\frac{\lambda+2}2(T^{(\mathcal{K})0}_{\hspace{0.6cm}0}+3\mu),
\end{equation}
where $\lambda$ and $\mu$ are undetermined constants. Using (\ref{46}) and (\ref{72}), we obtain
\begin{equation}\label{73}
T^{(\mathcal{K})0}_{\hspace{0.6cm}0}=\begin{cases}-3\Lambda,\\\frac{S(u)}{r^{\lambda+2}}-3\mu,\end{cases}\quad \text{for} \begin{array}{c}\lambda=-2,\\\lambda\neq -2,\end{array}
\end{equation}
where $\Lambda$ and $S(\mu)$ are integral constants. Substituting (\ref{73}) into (\ref{43}), we have
\begin{equation}
r(b^2)'+(b^2-1)=\begin{cases}3m^2\Lambda r^2,\\-\frac{m^2S(u)}{r^{\lambda}}+3m^2\mu r^2,\end{cases}\quad \text{for} \begin{array}{c}\lambda=-2,\\\lambda\neq -2,\end{array}
\end{equation}
and subsequently the solution as follows
\begin{equation}\label{75}
b^2=\begin{cases}1-\frac{r_s(u)}r+m^2\Lambda r^2,\\1-\frac{r_s(u)}r+\frac{m^2S(u)\ln r}r+m^2\mu r^ 2,\\1-\frac{r_s(u)}r+\frac{m^2S(u)}{(\lambda-1)r^\lambda}+m^2\mu r^2,\end{cases}\quad \text{for} \begin{array}{c} \lambda=-2,\\ \lambda=1,\\ \lambda\neq 1,-2.\end{array}
\end{equation}
In the case of $\lambda=-2$, the resulting solution is corresponding to generalized Vaidya solution, and the case of $\lambda\neq-2$ correlates closely with the solution of extended furry black hole \cite{Li2016b}, as we will see in the following.

\subsection{Generalized Vaidya solution}

For the $h(u,r)=\pm\left(\frac 2{3\alpha_3}+1\right)(u+r)$ and $S(u)=0$, we have
\begin{equation}
T^{(\mathcal{K})0}_{\hspace{0.6cm}0}=T^{(\mathcal{K})2}_{\hspace{0.6cm}2}=-\frac{48\alpha_3^2-64\alpha_4}{27\alpha_3^4},
\end{equation}
which corresponds to the case of $\lambda=-2$. Substituting (\ref{75}) into (\ref{20}) and (\ref{21}), we have the Ricci scalar $R=12m^2\Lambda$ and the Einstein tensor
\begin{equation}
G_{\mu\nu}=-\left(\frac{\dot{r}_s(u)}{r^2}+3m^2\Lambda b^2\right)\delta^0_\mu\delta^0_\nu.
\end{equation}
As a result, we have an expression of $r_s(u)$,
\begin{equation}
q(u)=\frac{dr_s(u)}{du}.
\end{equation}

Finally, the generalized Vaidya solution can be written as
\begin{equation}\label{79}
\begin{split}
b^2&=1-\frac{r_s(u)}r+\frac{(48\alpha_3^2-64\alpha_4)m^2r^2}{81\alpha_3^4},\\
\phi^0&=\pm \left(\frac 2{3\alpha_3}+1\right)(u+r),\\
\phi^i&=x^i.
\end{split}
\end{equation}

\subsection{Extended furry black holes}

For the case of $\dot{h}=0$ and $S(u)=S$, $r_s(u)=r_s$ ( $S$ and $r_s$ are constants), (\ref{75}) is rewritten as
\begin{equation}\label{80}
b^2=\begin{cases}1-\frac{r_s}r+\frac{m^2S\ln r}r+m^2\mu r^ 2,\\1-\frac{r_s}r+\frac{m^2S}{(\lambda-1)r^\lambda}+m^2\mu r^2,\end{cases}\quad \text{for} \begin{array}{c}  \lambda=1,\\ \lambda\neq 1,-2.\end{array}
\end{equation}
From (\ref{73}), we obtain the equation of $h(u,r)$ as follows
\begin{equation}
\begin{split}
&\left(3-\frac{16\alpha_4}{3\alpha_3^2}\right)\left(\left(\left(\frac 2{3\alpha_3}
+1\right)^2-h'^2\right)b^2\right)^{\frac 1 2}\\
&+\frac 4{9\alpha_3^2}(12\alpha_4-1)-3
=\frac S{r^{\lambda+2}}-3\mu.
\end{split}
\end{equation}
Therefore, we have
\begin{equation}
h=\pm\int\left(\left(\frac 2{3\alpha_3}+1\right)^2-\frac{S^2}{\left(3-\frac{16\alpha_4}{3\alpha_3}\right)^2b^2r^{2\lambda+4}}\right)^{\frac 1 2}dr,
\end{equation}
and
\begin{equation}
\mu=\frac{27\alpha_3^2-48\alpha_4+4}{27\alpha_3^2}.
\end{equation}
This solution is also free of singularities except for the conventional one appearing in GR. In fact, a straightforward calculation then shows that $I^{ab}$ are given by
\begin{equation}\label{84}
\begin{split}
I^{00}&=-b^2\left(\frac 2{3\alpha_3}+1\right)^2-\frac{S^2}{\left(3-\frac{16\alpha_4}{3\alpha_3}\right)^2r^{2\lambda+4}},\\
I^{0i}&=\mp b^2\left(\frac 2{3\alpha_3}+1\right)\left(\left(\frac 2{3\alpha_3}+1\right)^2+\frac{S^2}{\left(3-\frac{16\alpha_4}{3\alpha_3}\right)^2b^2r^{2\lambda+4}}\right)^{\frac 1 2}n^i,\\
I^{ij}&=\left(\frac 2{3\alpha_3}+1\right)^2((1-b^2)n^in^j-\delta^{ij}).
\end{split}
\end{equation}
Using the coordinate transformation (\ref{14}), we obtain the furry black hole solutions in the Schwarzschild coordinate: (i) for the case of $\lambda=1$,
\begin{equation}
\begin{split}
ds^2&=\left(1-\frac{r_s}r+\frac{m^2S\ln r}r+m^2\mu r^2\right)dt^2\\
&-\left(1-\frac{r_s}r+\frac{m^2S\ln r}r+m^2\mu r^2\right)^{-1}dr^2-r^2d\Omega^2,\\
\phi^0&=\pm\int\frac{\left[\left(3+\frac 2{\alpha_3}-\frac{16\alpha_4}{3\alpha_3}-\frac{32\alpha_4}{9\alpha_3^2}\right)b^2r^6-S^2\right]^{\frac 1 2}}{\left(3-\frac{16\alpha_4}{3\alpha_3}\right)br^3}dr,\\
\phi^i&=\left(\frac 2 {3\alpha_3}+1\right)x^i;
\end{split}
\end{equation}
and (ii) for the case of $\lambda\neq 1,-2$,
\begin{equation}
\begin{split}
ds^2&=\left(1-\frac{r_s}r+\frac{m^2S}{(\lambda-1)r^\lambda}+m^2\mu r^2\right)dt^2\\
&-\left(1-\frac{r_s}r+\frac{m^2S}{(\lambda-1)r^\lambda}+m^2\mu r^2\right)^{-1}dr^2-r^2d\Omega^2,\\
\phi^0&=\pm\int\frac{\left[\left(3+\frac 2{\alpha_3}-\frac{16\alpha_4}{3\alpha_3}-\frac{32\alpha_4}{9\alpha_3^2}\right)b^2r^{2\lambda+4}-S^2\right]^{\frac 1 2}}{\left(3-\frac{16\alpha_4}{3\alpha_3}\right)br^{\lambda+2}}dr,\\
\phi^i&=\left(\frac 2 {3\alpha_3}+1\right)x^i.
\end{split}
\end{equation}

\section{A furry Vaidya solution under the ansatz III}

In this section, we find out some generalized Vaidya solutions under self-consistent ansatz III in dRGT with $12\alpha_4=1+3\alpha_3+9\alpha_3^2$ and $\alpha_3\neq 0$. Let us suppose further
\begin{equation}\label{87}
h(u,r)=\frac{S}{r^\xi},
\end{equation}
where $\xi$ and $S$ are constants, then (\ref{32}) and (\ref{36}) can be reduced to
\begin{equation}\label{88}
T^{(\mathcal{K})0}_{\hspace{0.6cm}0}=\frac 1{(\xi+1)^2}\left(-\frac{2(\xi-1)S^2}{r^{2\xi+2}}+\frac{8\xi S}{r^{\xi+1}}+\xi^2-4\xi-1\right),
\end{equation}
and
\begin{equation}\label{89}
T^{(\mathcal{K})2}_{\hspace{0.6cm}2}=\frac 1{(\xi+1)^2}\left(-\frac{2\xi(\xi-1)S^2}{r^{2\xi+2}}-\frac{4\xi(\xi-1)S}{r^{\xi+1}}+\xi^2-4\xi-1\right).
\end{equation}

From (\ref{43}), (\ref{44}), (\ref{88}) and (\ref{89}), we obtain the exact solutions as follows
\begin{equation}\label{90}
\begin{split}
ds^2&=b^2(u,r)du^2+2dudr-r^2d\Omega^2,\\
\phi^0&=\frac{S}{r^\xi},\\
\phi^i&=\frac{S}{r^\xi}n^i,
\end{split}
\end{equation}
where $S$ and $\xi$ are constants and
\begin{widetext}
\begin{equation}\label{91}
b^2(u,r)-1+\frac{r_s(u)}r=\begin{cases}\frac{m^2}{(1+\xi)^2}\left(\frac{2(\xi-1)S^2}{(2\xi-1)r^{2\xi}}-\frac{8\xi S}{(\xi-2)r^{\xi-1}}+\frac{\xi^2-4\xi-1}3\right),\\\frac{m^2}{1+\xi}\left(\frac{2S^2}{9r^{2\xi}}+\frac{16S\ln r}{3r}-\frac 5 9\right),\\\frac{m^2}{1+\xi}\left(\frac{16S^2r^\xi}9+\frac{2S\ln r}{3r}-\frac {11}{18}\right),\\\end{cases}\quad \text{for} \begin{array}{c} \alpha_3\neq -\frac 2 9,-\frac 4 9,\\ \alpha_3=-\frac 2 9,\\ \alpha_3=-\frac 4 9,\end{array}
\end{equation}
\end{widetext}
and
\begin{equation}\label{92}
\xi=-\left(1+\frac 2{2\alpha_3}\right).
\end{equation}
As the generalized Vaidya solutions, there is still a relation $\dot{r}_s(u)=q(u)$. Obviously, if $\alpha_3\rightarrow \infty$ we have $\xi\rightarrow -1$; if $\alpha_3\rightarrow 0$, we have $\xi\rightarrow \infty$. If and only if $\alpha_3=-\frac{3\pm\sqrt{5}}6$, the solution (\ref{91}) is asymptotically flat. In the case of $q(u)=0$, we obtain new furry black holes from (\ref{90})-(\ref{92}).

\section{Conclusion and discussion}

In GR, the Vaidya solution is a nonstatic generalization of the Schwartzschild metric and has some unique features, and its massive deformation also plays an interesting role in dRGT. In this work, we have developed a study of the Vaidya solution and its generalization in dRGT if the St\"{u}ckeberg fields are taken as some self-consistent ansatz. Under the ansatz I, we obtain the Vaidya solution in dRGT. Under the ansatz II and III, we obtain the Vaidya-de Sitter and the furry Vaidya solution, respectively. As by-products, we obtain a series of the furry black holes.

The Vaidya solution and its generalization in dRGT massive gravity describe the  black holes with a variable horizon. For the metric
\begin{equation}\label{93}
ds^2=b^2(u,r)du^2+2dudr-r^2d\Omega^2,
\end{equation}
we take Schwarzschild coordinate
\begin{equation}
t=u+\int^r_{r_0}\frac{dr}{b^2(0,r)},
\end{equation}
then (\ref{93}) can be rewritten as
\begin{equation}
\begin{split}
ds^2=&b^2(u,r)dt^2-\frac{2\left(b^2(u,r)-b^2(0,r)\right)}{b^2(0,r)}dtdr\\
&-\frac{2b^2(0,r)-b^2(u,r)}{b^4(0,r)}dr^2-r^2d\Omega^2.
\end{split}
\end{equation}
There is an infinite red-shift surface in $b^2(u,r)=0$, which corresponds to the event horizon. Especially, the Vaidya solution (\ref{58}) can be rewritten as
\begin{equation}
\begin{split}
ds^2=&\left(1-\frac{r_s(u)}r\right)dt^2-\frac{2\left(r_s(0)-r_s(u)\right)}{r-r_s(0)}dtdr\\
&-\frac{r(r-2r_s(0)+r_s(u))}{(r-r_s(0))^2}dr^2-r^2d\Omega^2,
\end{split}
\end{equation}
so the event horizon is
\begin{equation}
r_s(u)=r_s(0)+\int^u_0q(u)du,
\end{equation}
and $r_s(u)$ is a variable horizon. In reality, the radius of generalized Vaidya solutions are changeable, not only event horizon but also cosmological one.

For all solutions, the singularities in the invariant $I^{ab}$ are absent. In fact, the invariant $I^{ab}$ can be explicitly expressed as
\begin{equation}
\begin{split}
I^{00}&=2\dot{h}h'-b^2h'^2,\\
I^{0i}&=(2\dot{h}-b^2h')\phi'n^i,\\
I^{ij}&=(2\dot{\phi}\phi'-b^2\phi'^2+\frac{\phi^2}{r^2})n^in^j-\frac{\phi^2}{r^2}\delta^{ij},
\end{split}
\end{equation}
where $(n^1,n^2,n^3)=(\sin\theta\cos\phi,\sin\theta\sin\phi,\cos\theta)$. Obviously, the singularities in the invariant $I^{ab}$ are absent except for the physical singularity $r=0$ in GR, so that these solutions of massive gravity may be regarded as candidates for the black hole in dRGT.

In addition, one may be anxious that the scalar perturbations on these backgrounds are infinitely strongly coupled in light of the result of Ref. \cite{Rham2011b}. It is found that the de Sitter background has infinitely strongly coupled fluctuations in the decoupling limit for the parameters chosen as $9\alpha_3^2+3\alpha_3-12\alpha_4+1=0$\cite{Rham2011b}. Under the ansatz I and II, we have $\pi^0=-h(r)$ and $\pi^i=(1-\beta)x^i$ for the furry black holes. From (\ref{9}), we obtain the vector mode $A^0=-\frac{\Lambda^3}mh(r)$ and $A^i=0$ which are different from those studied in \cite{Rham2011b}. However, we have to meet this question under the ansatz III unless we consider asymptotically flat background. Finally, we can also discuss the Kerr solution using our method developed in this work and will do so in a forthcoming paper.

\acknowledgments
This work is supported by National Science Foundation of China grant No.~10671128 and Key Project of Chinese Ministry of Education grant, No.~211059.

\bibliography{ref}

\begin{thebibliography}{17}
\expandafter\ifx\csname natexlab\endcsname\relax\def\natexlab#1{#1}\fi
\expandafter\ifx\csname bibnamefont\endcsname\relax
  \def\bibnamefont#1{#1}\fi
\expandafter\ifx\csname bibfnamefont\endcsname\relax
  \def\bibfnamefont#1{#1}\fi
\expandafter\ifx\csname citenamefont\endcsname\relax
  \def\citenamefont#1{#1}\fi
\expandafter\ifx\csname url\endcsname\relax
  \def\url#1{\texttt{#1}}\fi
\expandafter\ifx\csname urlprefix\endcsname\relax\def\urlprefix{URL }\fi
\providecommand{\bibinfo}[2]{#2}
\providecommand{\eprint}[2][]{\url{#2}}

\bibitem[{\citenamefont{de~Rham}(2014)}]{Rham2014}
\bibinfo{author}{\bibfnamefont{C.}~\bibnamefont{de~Rham}},
  \bibinfo{journal}{Living Rev. Relativity} \textbf{\bibinfo{volume}{17}},
  \bibinfo{pages}{7} (\bibinfo{year}{2014}).

\bibitem[{\citenamefont{Boulware and Deser}(1972)}]{Boulware1972}
\bibinfo{author}{\bibfnamefont{D.~G.} \bibnamefont{Boulware}} \bibnamefont{and}
  \bibinfo{author}{\bibfnamefont{S.}~\bibnamefont{Deser}},
  \bibinfo{journal}{Phys. Rev. D} \textbf{\bibinfo{volume}{6}},
  \bibinfo{pages}{3368} (\bibinfo{year}{1972}).

\bibitem[{\citenamefont{Gabadadze}(2009)}]{Gabadadze2009}
\bibinfo{author}{\bibfnamefont{G.}~\bibnamefont{Gabadadze}},
  \bibinfo{journal}{Phys. Lett. B} \textbf{\bibinfo{volume}{681}},
  \bibinfo{pages}{89} (\bibinfo{year}{2009}).

\bibitem[{\citenamefont{de~Rham}(2010)}]{Rham2010a}
\bibinfo{author}{\bibfnamefont{C.}~\bibnamefont{de~Rham}},
  \bibinfo{journal}{Phys. Lett. B} \textbf{\bibinfo{volume}{688}},
  \bibinfo{pages}{137} (\bibinfo{year}{2010}).

\bibitem[{\citenamefont{de~Rham and Gabadadze}(2010)}]{Rham2010b}
\bibinfo{author}{\bibfnamefont{C.}~\bibnamefont{de~Rham}} \bibnamefont{and}
  \bibinfo{author}{\bibfnamefont{G.}~\bibnamefont{Gabadadze}},
  \bibinfo{journal}{Phys. Rev. D} \textbf{\bibinfo{volume}{82}},
  \bibinfo{pages}{044020} (\bibinfo{year}{2010}).

\bibitem[{\citenamefont{de~Rham
  et~al.}(2011{\natexlab{a}})\citenamefont{de~Rham, Gabadadze, and
  Tolley}}]{Rham2011a}
\bibinfo{author}{\bibfnamefont{C.}~\bibnamefont{de~Rham}},
  \bibinfo{author}{\bibfnamefont{G.}~\bibnamefont{Gabadadze}},
  \bibnamefont{and} \bibinfo{author}{\bibfnamefont{A.~J.}
  \bibnamefont{Tolley}}, \bibinfo{journal}{Phys. Rev. Lett.}
  \textbf{\bibinfo{volume}{106}}, \bibinfo{pages}{231101}
  (\bibinfo{year}{2011}{\natexlab{a}}).

\bibitem[{\citenamefont{Zhang and Li}(2016)}]{Zhang2016}
\bibinfo{author}{\bibfnamefont{H.}~\bibnamefont{Zhang}} \bibnamefont{and}
  \bibinfo{author}{\bibfnamefont{X.~Z.} \bibnamefont{Li}},
  \bibinfo{journal}{Phys. Rev. D} \textbf{\bibinfo{volume}{93}},
  \bibinfo{pages}{124039} (\bibinfo{year}{2016}).

\bibitem[{\citenamefont{Li et~al.}(2016{\natexlab{a}})\citenamefont{Li, Li, and
  Xi}}]{Li2016a}
\bibinfo{author}{\bibfnamefont{P.}~\bibnamefont{Li}},
  \bibinfo{author}{\bibfnamefont{X.~Z.} \bibnamefont{Li}}, \bibnamefont{and}
  \bibinfo{author}{\bibfnamefont{P.}~\bibnamefont{Xi}},
  \bibinfo{journal}{Class. Quantum Grav.} \textbf{\bibinfo{volume}{33}},
  \bibinfo{pages}{115004} (\bibinfo{year}{2016}{\natexlab{a}}).

\bibitem[{\citenamefont{Li et~al.}(2016{\natexlab{b}})\citenamefont{Li, Li, and
  Xi}}]{Li2016b}
\bibinfo{author}{\bibfnamefont{P.}~\bibnamefont{Li}},
  \bibinfo{author}{\bibfnamefont{X.~Z.} \bibnamefont{Li}}, \bibnamefont{and}
  \bibinfo{author}{\bibfnamefont{P.}~\bibnamefont{Xi}}, \bibinfo{journal}{Phys.
  Rev. D} \textbf{\bibinfo{volume}{93}}, \bibinfo{pages}{064040}
  (\bibinfo{year}{2016}{\natexlab{b}}).

\bibitem[{\citenamefont{Hinterbichler}(2012)}]{Hinterbichler2012}
\bibinfo{author}{\bibfnamefont{K.}~\bibnamefont{Hinterbichler}},
  \bibinfo{journal}{Rev. Mod. Phys.} \textbf{\bibinfo{volume}{84}},
  \bibinfo{pages}{671} (\bibinfo{year}{2012}).

\bibitem[{\citenamefont{Berezhiani et~al.}(2012)\citenamefont{Berezhiani,
  Chkareuli, de~Rham, Gabadadze, and Tolley}}]{Berezhiani2012}
\bibinfo{author}{\bibfnamefont{L.}~\bibnamefont{Berezhiani}},
  \bibinfo{author}{\bibfnamefont{G.}~\bibnamefont{Chkareuli}},
  \bibinfo{author}{\bibfnamefont{C.}~\bibnamefont{de~Rham}},
  \bibinfo{author}{\bibfnamefont{G.}~\bibnamefont{Gabadadze}},
  \bibnamefont{and} \bibinfo{author}{\bibfnamefont{A.~J.}
  \bibnamefont{Tolley}}, \bibinfo{journal}{Phys. Rev. D}
  \textbf{\bibinfo{volume}{85}}, \bibinfo{pages}{044024}
  (\bibinfo{year}{2012}).

\bibitem[{\citenamefont{Vainshtein}(1972)}]{Vainshtein1972}
\bibinfo{author}{\bibfnamefont{A.~I.} \bibnamefont{Vainshtein}},
  \bibinfo{journal}{Phys. Lett. B} \textbf{\bibinfo{volume}{39}},
  \bibinfo{pages}{393} (\bibinfo{year}{1972}).

\bibitem[{\citenamefont{Vaidya}(1943)}]{Vaidya1943}
\bibinfo{author}{\bibfnamefont{P.~C.} \bibnamefont{Vaidya}},
  \bibinfo{journal}{Curr. Sci.} \textbf{\bibinfo{volume}{12}},
  \bibinfo{pages}{183} (\bibinfo{year}{1943}).

\bibitem[{\citenamefont{Vaidya}(1953)}]{Vaidya1953}
\bibinfo{author}{\bibfnamefont{P.~C.} \bibnamefont{Vaidya}},
  \bibinfo{journal}{Nature} \textbf{\bibinfo{volume}{171}},
  \bibinfo{pages}{260} (\bibinfo{year}{1953}).

\bibitem[{\citenamefont{Arkani-Hamed et~al.}(2003)\citenamefont{Arkani-Hamed,
  Georgi, and Schwartz}}]{Arkani2003}
\bibinfo{author}{\bibfnamefont{N.}~\bibnamefont{Arkani-Hamed}},
  \bibinfo{author}{\bibfnamefont{H.}~\bibnamefont{Georgi}}, \bibnamefont{and}
  \bibinfo{author}{\bibfnamefont{M.~D.} \bibnamefont{Schwartz}},
  \bibinfo{journal}{Ann. Phys. (Amsterdam)} \textbf{\bibinfo{volume}{305}},
  \bibinfo{pages}{96} (\bibinfo{year}{2003}).

\bibitem[{\citenamefont{Carmeli and Kaye}(1977)}]{Carmeli1977}
\bibinfo{author}{\bibfnamefont{M.}~\bibnamefont{Carmeli}} \bibnamefont{and}
  \bibinfo{author}{\bibfnamefont{M.}~\bibnamefont{Kaye}},
  \bibinfo{journal}{Ann. Phys. (N. Y. )} \textbf{\bibinfo{volume}{103}},
  \bibinfo{pages}{97} (\bibinfo{year}{1977}).

\bibitem[{\citenamefont{de~Rham
  et~al.}(2011{\natexlab{b}})\citenamefont{de~Rham, Gabadadze, Heisenberg, and
  Pirtskhalava}}]{Rham2011b}
\bibinfo{author}{\bibfnamefont{C.}~\bibnamefont{de~Rham}},
  \bibinfo{author}{\bibfnamefont{G.}~\bibnamefont{Gabadadze}},
  \bibinfo{author}{\bibfnamefont{L.}~\bibnamefont{Heisenberg}},
  \bibnamefont{and}
  \bibinfo{author}{\bibfnamefont{D.}~\bibnamefont{Pirtskhalava}},
  \bibinfo{journal}{Phys. Rev. D} \textbf{\bibinfo{volume}{83}},
  \bibinfo{pages}{103516} (\bibinfo{year}{2011}{\natexlab{b}}).

\end{thebibliography}
\bibliographystyle{apsrev}
\end{document}